\documentclass[reprint,superscriptaddress,nofootinbib,amsmath,amssymb,aps,preprintnumbers]{revtex4-2}

\usepackage{graphicx}
\usepackage{dcolumn}
\usepackage{bm}
\usepackage{hyperref}
\hypersetup{
    colorlinks,
    citecolor=blue,
    linkcolor=blue,
    urlcolor=blue,
}

\def\planck{E_\text{Planck}}
\def\nplanck{1.22\times{}10^{19}~\text{GeV}}
\def\gev{\text{GeV}}
\def\tev{\text{TeV}}

\def\textd{\text{d}}
\def\eliv{E_\text{LIV}}

\newcommand{\textadded}[1]{#1}

\begin{document}

\preprint{H.~Li \& B.-Q.~Ma, \href{https://doi.org/10.1088/1475-7516/2023/10/061}{JCAP10(2023)061} 
[\href{https://arxiv.org/abs/2306.02962}{arXiv:2306.02962}] }

\title{Revisiting Lorentz invariance violation from GRB 221009A}

\author{Hao Li}
 \email{haolee@pku.edu.cn}
\affiliation{School of Physics, 
Peking University, Beijing 100871, China}

\author{Bo-Qiang Ma}%
 \email{mabq@pku.edu.cn}
\affiliation{School of Physics,
Peking University, Beijing 100871, China}
\affiliation{Center for High Energy Physics, Peking University, Beijing 100871, China}%
\affiliation{Collaborative Innovation Center of Quantum Matter, Beijing, China}

\begin{abstract}
As a potential consequence of Lorentz invariance violation~(LIV), threshold anomalies open a window to study LIV\@.
Recently the Large High Altitude Air Shower Observatory~(LHAASO) reported that more than 5000 photons from GRB 221009A have been observed with energies above 500~GeV and up to $18~\text{TeV}$.
In the literature, it is suggested that this observation  
may have tension with
the standard model result because extragalactic background light~(EBL) can prevent photons around 18~TeV from reaching the earth and that LIV induced threshold anomalies might be able
to explain the observation. 
In this work we further study this proposal with more detailed numerical calculation for different LIV scales and redshifts of the sources.
We find that GRB 221009A is a rather unique opportunity to search LIV, and a LIV scale $E_\text{LIV} \lesssim E_\text{Planck}\approx 1.22\times 10^{19}~\text{GeV}$ is 
feasible 
to the observation of GRB 221009A on 9 October, 2022.
\end{abstract}

\keywords{Lorentz invariance violation, extragalactic background light, threshold anomaly, gamma-ray burst}

\maketitle

\section{Introduction\label{introduction}}

Lorentz symmetry, one of the cornerstones of modern physics, although having been proven to be satisfied to very high accuracy, is continuously challenged after the proposal of searching for Lorentz invariance violation (LIV) through high energy photons from astrophysical objects such as gamma-ray bursts about thirty years ago~\cite{Amelino-Camelia1997a,Amelino-Camelia1998}.
There are several constructions of theories of quantum gravity~(QG) suggesting some LIV phenomena as the low-energy remnants of QG effects at the Planck scale~$\planck{}\approx\nplanck{}$~\cite{Amelino-Camelia1997a,Ellis1999,Ellis2000,Ellis2004,Ellis2008,Li2009,Li:2021pre,LI2021104380,Gambini1999,Alfaro2002a,Alfaro2002b,LI2023,AMELINO-CAMELIA2002a,Magueijo2002,AMELINO-CAMELIA2002b, Amelino-Camelia2005,Kowalski-Glikman:2004fsz}.
Since its tight relationship to our understanding of QG, pursuing the evidence of LIV or falsifying the proposal of the existence of LIV is becoming more important and urgent.

Recently a unique opportunity to search LIV appears: the extraordinarily bright gamma-ray burst~(GRB) GRB 221009A~\cite{GBM1,GBM2,LAT1,LAT2,Swift1,Swift2,Swift3} located at redshift $z=0.1505$~\cite{Redshift1,Redshift2} was observed in the energy range from 500~GeV to about 18~TeV by the Large High Altitude Air Shower Observatory~(LHAASO), with more than 5000 photons recorded~\cite{LHAASO}.
However this result seems to be an alarm that we may need new physics to interpret it since we expect that photons around 18~TeV are unlikely observable 
due to 
the attenuation of extragalactic background light~(EBL)~\cite{Li:2022vgq}.
It immediately triggered the studies of whether LIV could come to the rescue~\cite{Li:2022wxc,Baktash:2022gnf,Finke:2022swf,Zhu_2023}.
The key point roots in the possibility of LIV induced threshold anomalies, which influence the attenuation behaviours of such energetic photons in the universe so that the excess beyond the standard results based on Lorentz invariant theories is allowed.
In other words, energetic photons, which should be absorbed by background light in the universe and thus can hardly be observed now have more chances to reach the earth as a consequence of LIV induced threshold anomalies.
\textadded{%
Of course it should be pointed out that observing the excess of high energy photons cannot be a conclusive evidence for LIV, since there exist other explanations.
Indeed, in the literature, several suggestions concerning non-standard mechanisms similar to LIV are widely discussed, including utilizing axion-like particles~(ALPs)~\cite{Galanti:2022pbg,Baktash:2022gnf,Troitsky:2022xso,Nakagawa:2022wwm,Zhang:2022zbm,Wang:2023okw} and heavy/sterile neutrinos~\cite{Cheung:2022luv,Smirnov:2022suv,Brdar:2022rhc,Huang:2022udc,Guo:2023bpo} to understand the observations.
Meanwhile it is also argued that even within the framework of standard physics, room for understanding the observations of LHAASO exists and one possibility is that the above 10~TeV photons observed by LHAASO are basically secondaries produced by interactions of ultra-high-energy cosmic rays~(UHECRs) on their path to Earth~\cite{AlvesBatista:2022kpg,Das:2022gon,Mirabal:2022ipw}.
At present none of these options is ruled out, and we would like to discuss this topic in more detail in Sec.~\ref{discussions}.
Nevertheless there remains room for LIV as a candidate for understanding the unusual observations of around 18~TeV photons from GRB 221009A\@\footnote{\textadded{At present we can not exclude the possibility that these around
18~TeV photons are not from GRB 221009A, and we hope in the
future LHAASO will provide more information on the correlation of the 18 TeV photon event with GRB 221009A.}}.
Therefore, in this work we provide more details of understanding the observations of GRB 221009A using LIV and hope that this work will serve as a preparation and a starting point for future studies.
}
In the following, at first, we introduce the basic background of this kind of anomalies and its impact on photon attenuation by the background light in the universe. As for the photons reported by LHAASO, it is sufficient to only consider the background light component dubbed extragalactic background light~(EBL), other sources like cosmic microwave background~(CMB) would not be taken into consideration in this work.

\section{Theoretical background\label{background}}

In this section, we briefly introduce the theoretical 
backgrounds 
for the analyses of this work, including threshold anomalies in LIV and EBL attenuation of high energy photons in special relativity~(SR) and in the LIV framework.
More details about LIV can be found in the reviews~\cite{Mattingly2005,Amelino-Camelia2013,He:2022gyk} and references therein.
For EBL one can refer to Refs.~\cite{Dominguez2011,Biteau2015} and references therein.

\subsection{Modified dispersion relations and threshold anomalies in LIV\label{mdrandta}}

An intensively studied feature of LIV is the modified dispersion relations~(MDRs) that differ from the SR ones by terms 
depending 
on the ratio $\left(E/\planck{}\right)$, where $E$ represents the particle energy.
For photons, one can adopt a model-independent form of the MDR to the leading order of $\left(E/\planck{}\right)$\footnote{For our purpose, it suffices to study only the leading order LIV deviation of the MDR, while the complete knowledge of the MDR is not quite clear in general, 
except for those in the doubly special relativity~(DSR) models~\cite{AMELINO-CAMELIA2002a,Magueijo2002,AMELINO-CAMELIA2002b, Amelino-Camelia2005,Kowalski-Glikman:2004fsz}.}:
\begin{equation}
    \label{generalmdr}
    E^2 = p^2\left(1-s{\left(\frac{p}{E_\text{LIV,n}}\right)}^n\right),
\end{equation}
where $s=\pm 1$ represents the possible helicity dependence, $n=1,2,3,\ldots$, and $E_\text{LIV,n}$, the scale of the onset of LIV, is of or close to the order of the magnitude of $\planck{}$.
It is clear from Eq.~\eqref{generalmdr} that if $s=+1$, the LIV term plays the role of an effective (running) mass, and if $s=-1$ photons maybe superluminal but could decay rapidly through the process $\gamma\to e^-e^+$~\cite{Li:2021pre2}, so we only focus on the subluminal case $s=+1$.
If we further assume that the energy-momentum conservation law 
still holds, then the threshold of a process in which photons take part could be anomalous~\cite{Mattingly2003, Jacobson2003,LI2021a}.
Indeed, let us consider two photons with four momenta $p_1=(E,0,0,p)$, where $E$ and $p$ obey Eq.~\eqref{generalmdr}, and $p_2=(\varepsilon_b,0,0,-\varepsilon_b)$ respectively, 
and we may imagine that the photon with $p_1$ is the high energy one from GRB, while $p_2$ is from EBL thus $\varepsilon_b\ll E$ and we can neglect the LIV of this photon.
This two photons may produce an electron and a positron, then the ordinary energy-momentum conservation law states that:
\begin{equation}
    \label{conservationlaw}
    4m_e^2 = (p_1+p_2){}^2,
\end{equation}
and it is noteworthy that we do not add LIV terms for electrons (or positrons) for both experimental and theoretical considerations~\cite[see][for example]{Li:2021pre,LI2021104380,Li:2022ugz}.
It is straight forward to show that Eq.~\eqref{conservationlaw} leads to~\cite{LI2021a}
\begin{equation}
    \label{xi}
    \xi_n = \frac{4\varepsilon_b}{p^{n-1}}-\frac{4m_e^2}{p^n},\ \text{for } p>0,
\end{equation}
to the leading order, where $\xi_n^{-1}:=E^n_\text{LIV,n}$ for convenience.
One can easily verify Eq.~\eqref{xi} by restoring the SR threshold for photons by setting $\xi_n=+\infty$, obtaining
\begin{equation}
    \label{srthreshold}
    E_{\mathrm{th}}^\text{SR}=\frac{m_e^2}{\varepsilon_b},
\end{equation}
with which we could calculate the typical scales for CMB attenuation and EBL attenuation as about $411~\tev{}$ and $261~\gev{}$ respectively~\cite{LI2021a,Li:2022vgq}, and it is obvious that our focus on EBL is reasonable from these results.
There exist many analyses about Eq.~\eqref{xi} in the literature~\cite{Mattingly2003,Jacobson2003,LI2021a}, and for this work we concentrate on the $n=1$ case so that we rewrite $\xi=\xi_1$.
The relevant results utilized in this work then are summarized as follows~\cite{LI2021a}:
\begin{description}
  \item[Case I] If $\xi>\xi_c:={16\times\varepsilon_b^3}/{(27m_e^4)}$, subluminal photons cannot be absorbed by background photons with energy $\varepsilon_b$ 
  because that the process $\gamma\gamma\to e^- e^+$ is forbidden kinematically.
  \item[Case II] If $0<\xi<\xi_c$, a subluminal photon can be absorbed only when its energy falls into a certain closed interval with its lower bound greater than $E_\text{th}$. 
  There is also an upper threshold and the $\varepsilon_b$ background is again transparent to photons with energy exceeding this upper threshold.
\end{description}
Of course a non-$\pi{}$ angle $\theta{}$ between the two photons involed in the process can raise the threshold, but the reactions are still suppressed because the pair-production process with certain permissible configurations in the standard case are not allowed to occur due to  LIV, and therefore we may find excesses in the spectra of GRB photons.
This picture provides the starting point of threshold anomaly studies including this work.

\subsection{EBL attenuation without and with LIV\label{ebl}}

The distribution of EBL still needs to be constrained further, however we adopt a plausible model~\cite{Dominguez2011} which possesses most of the common properties of existing models, and extension to other models is straightforward.

For very high energy photons, there are two major processes that contribute to the absorption by background light: the pair-production process $\gamma\gamma\to e^-e^+$ and the double pair-production process $\gamma\gamma\to e^-e^-e^+e^+$~\cite{Ruffini2016}.
The process dominating the attenuation of 500~GeV to 18~TeV photons is the first one, i.e., the pair-production process~\cite{Ruffini2016}.
For the pair-production process, the cross-section is readily obtained from quantum electrodynamics~(QED)~\cite{Nikishov1962,Gould1967b}:
\begin{eqnarray}
    \sigma_{\gamma\gamma}(\beta_0) &= &\frac{3\sigma_T}{16}(1-\beta_0^2) \nonumber \\
    & & \times\left[2\beta_0(\beta_0^2-2)+(3-\beta_0^4)\ln\frac{1+\beta_0}{1-\beta_0}\right],\label{crosssection}
\end{eqnarray}
where $\sigma_T$ is the Thomson cross-section, and
\begin{equation}
    \label{beta0}
    \beta_0 = \sqrt{1-\frac{2m_e^2}{E\varepsilon}\frac{1}{1-\mu}},
\end{equation}
with $\cos\theta{}$ defined to be $\mu{}$.
As for this work, the energies in Eq.~\eqref{beta0} are the present-day measurements of a GRB photon and an EBL photon.
It is also worth noting that Eq.~\eqref{beta0} automatically manifests the threshold by ensuring $\beta$ belonging to the real number field $\mathbb{R}$:
\begin{equation}
    E\ge \frac{m_e^2}{\varepsilon}\frac{2}{1-\cos\theta},\label{threshold}
\end{equation}
To use the cross-section~\eqref{crosssection} in the calculation of EBL attenuation, we should further take into consideration the expansion of the universe, and hence $E$ and $\varepsilon{}$ should be multiplied by $(1+z)$.
Thus for convenience we define
\begin{equation}
    \beta(E,z,\varepsilon,\mu)=\sqrt{1-\frac{2m_e^2}{E\varepsilon}\frac{1}{1-\mu}\left(\frac{1}{1+z}\right){}^2},\label{beta}
\end{equation}
and clearly $\beta_0=\beta(E,0,\varepsilon,\mu)$.
With this cross-section we can calculate the optical depth\footnote{The integration over energy is understood to be performed above the threshold.}:
\begin{widetext}
\begin{equation}
    \label{opticaldepth}
    \tau_{\gamma\gamma}(E,z)=\int^z_0\textd z^\prime \frac{\partial l}{\partial z^\prime}(z^\prime)\int_0^\infty \textd \varepsilon\frac{\partial n}{\partial\varepsilon}(\varepsilon,z^\prime)\int_{-1}^{1}\textd \mu\frac{1-\mu}{2}\sigma_{\gamma\gamma}\left[\beta(E,z^\prime,\varepsilon,\mu)\right],
\end{equation}
\end{widetext}
where $\partial n(\varepsilon,z^\prime)/\partial\varepsilon{}$ is the number density of EBL photons of energy $\varepsilon{}$ at redshift $z^\prime{}$, and
\begin{equation}
    \frac{\partial l}{\partial z^\prime} = \frac{1}{H_0}\frac{1}{1+z^\prime}\frac{1}{\sqrt{\Omega_\Lambda+\Omega_M(1+z^\prime){}^3}},
\end{equation}
is obtained from the Friedmann-Robertson-Walker cosmology with the Hubble constant $H_0=70~\text{km}\text{s}^{-1}\text{Mpc}^{-1}$ as well as the cosmological parameters $\Omega_\Lambda=0.7$ and $\Omega_M=0.3$.
Then the intrinsic and observed fluxes (represented by $F_\text{int}$ and $F_\text{obs}$ respectively) can be related by
\begin{equation}
    F_\text{obs}=F_\text{int}\times e^{-\tau_{\gamma\gamma}}.\label{flux}
\end{equation}

However, once LIV takes effect in the pair-production process, the modified optical depth $\tau_\text{LIV}$ may be different from $\tau_{\gamma\gamma}$.
The deviation is at least two-fold.
First, the topic of this work, the threshold anomaly of the pair-production process~\eqref{xi}, makes the integration over $\varepsilon{}$ in Eq.~\eqref{opticaldepth} change~\cite{Kifune:1999ex,Protheroe:2000hp}, i.e., the phase space of particle kinematics in the LIV case is different from that in the standard SR case.
Second, it is natural that one assumes that the functional form of the cross-section of the pair-production process, $\sigma_{\gamma\gamma}$, gets modified for reasons like extra terms in the corresponding Lagrangian.
We do not have enough knowledge of the modified cross-section since this needs the complete understanding of the LIV dynamics, but we can expect that at least in the low energy limit:
\begin{eqnarray}
    \sigma^\text{LIV}_{\gamma\gamma}(E,\cdots) &=& \sigma_{\gamma\gamma}(E,\cdots) \nonumber \\
    & & \times\left(1+\vartheta\left(\frac{E}{E_\text{LIV}}\right)+\cdots\right),\label{livcrosssection}
\end{eqnarray}
where $\vartheta{}$ is a constant.
Of course, there may exist other kinds of corrections to the calculation of $\tau_\text{LIV}$, such as the modification of the cosmological model~(the function $\partial l/\partial z^\prime{}$) and the understanding of the EBL~(the density $n_\varepsilon:=\partial n/\partial\varepsilon{}$).
In the following we first focus on the first element, i.e.\ the threshold anomaly of the pair-production process, while in Sec.~\ref{discussions} we briefly go back to the second element and discuss the possible effects.
Other sources of corrections~($\partial l/\partial z^\prime{}$ and $n_\varepsilon{}$) are left for future studies.

\section{Analyses\label{analyses}}

In this section we perform the analyses of photon attenuation for photons of GRB 221009A observed by LHAASO~\cite{LHAASO}, and meanwhile we extend our analyses into other cases: a GRB at $z=0.5$ which is the typical distance of a short burst, and a GRB at $z=2.15$ that is the typical redshift of a long burst.
To make the numerical calculation feasible, we need the information of $n_\varepsilon{}$ and the exact form of $\sigma^\text{LIV}_{\gamma\gamma}$.
In this work we 
only adopt the EBL model~$n_\varepsilon{}$ of Dom{\'\i}nguez {\em et al.}~\cite{Dominguez2011}.
Indeed replacement with other models is quite straightforward, and the difference in the results is supposed to be mild, since the adopted model captures the major features of existing EBL models already.
However to know the exact form of $\sigma^\text{LIV}_{\gamma\gamma}$ we should understand the LIV dynamics,
which is still lacking. 
Alternatively in the calculation of the LIV cases we assume that $\sigma^\text{LIV}_{\gamma\gamma}=\sigma_{\gamma\gamma}$, and only consider the LIV effects brought by the change in the threshold.
The possible consequence of the modification of the cross-section is discussed in Sec.~\ref{discussions}.

\subsection{GRB 221009A\label{221009A}}

\begin{figure}[htbp]
    \centering
    \includegraphics[scale=0.53]{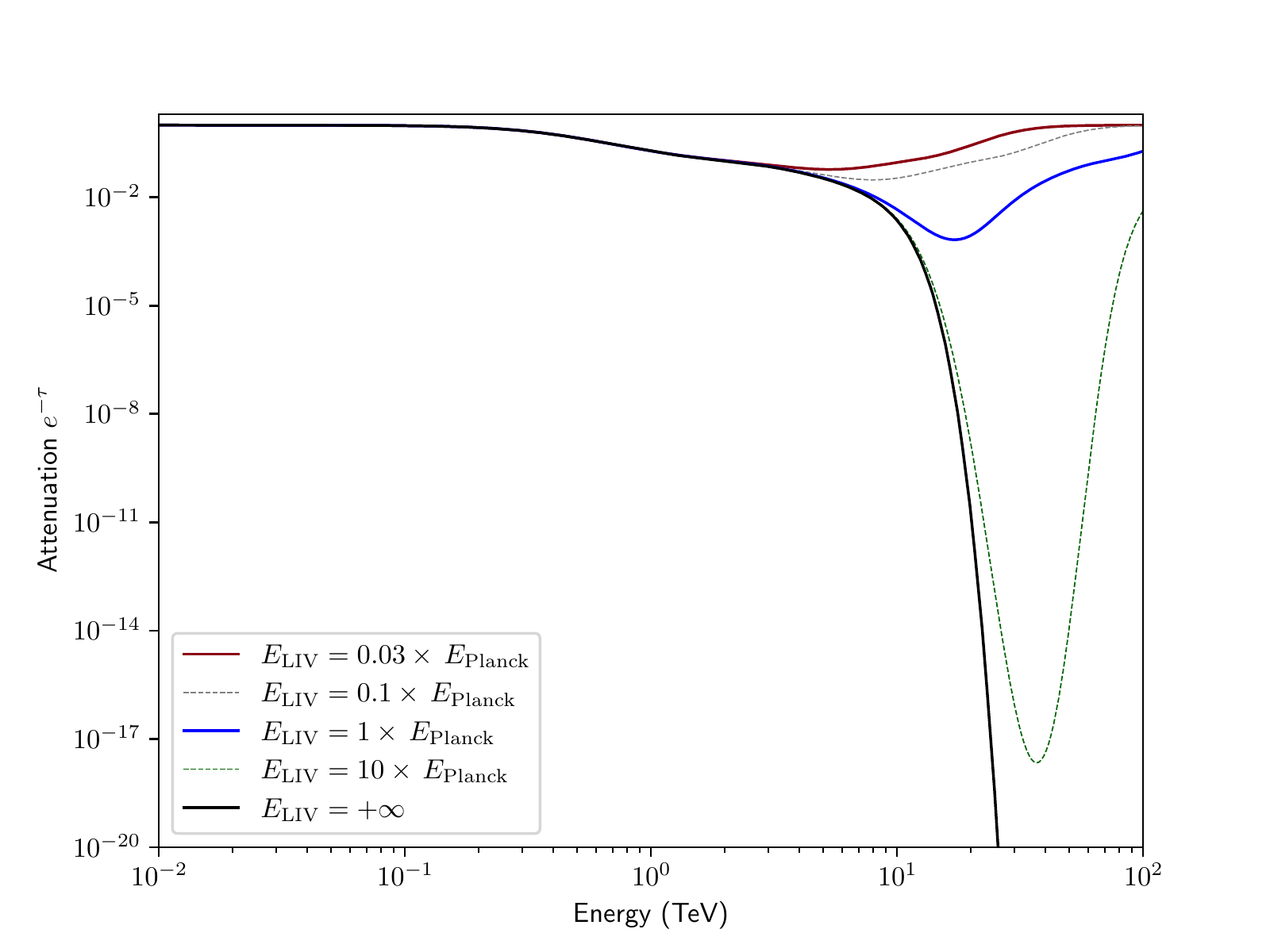}
    \caption{The EBL attenuation of $z=0.1505$ with different LIV scales $\eliv{}$~\cite{Dominguez2011,manuel-meyer}.\label{ebl_01505}}
\end{figure}

GRB 221009A is located at $z=0.1505$~\cite{Redshift1,Redshift2}, therefore we first draw the $E$-$e^{-\tau}$ plot of $z=0.1505$ in Fig.~\ref{ebl_01505} with different LIV scales utilizing {\bfseries ebltable}~\cite{manuel-meyer}.
The vertical axis of this figure can also be understood as the survival probability of photons from the GRB after being absorbed by EBL\@.
In the figure we choose the LIV scale $\eliv{}$ to be 
$3.6\times 10^{17}~\gev=0.03\times\planck$~\footnote{This is a phenomenological suggestion as a lower bound for LIV~\cite{Shao2010f, Xiao2009, Xiao2009a, Shao2010b, Zhang2015, Xu2016a, Xu2016,  Xu2018, Liu2018, ZHU2021136518, Li2020} which we list for comparison here.}, $\planck/10$, $\planck{}$, $10\times\planck{}$, and the non-LIV case which can also be considered as $\eliv=+\infty{}$.
From Fig.~\ref{ebl_01505} we can immediately infer some useful information.
First, from 500~GeV to around 10~TeV, the effects of different LIV scales are not distinguishable enough from the standard case.
Second, if $\eliv\gtrsim 10\times\planck{}$, the LIV effect is not remarkable for $z=0.1505$.
Third, we can conclude that so long as $\eliv\lesssim\planck{}$ the LIV effects become dramatic enough.
However we should point out that as we can see from Fig.~\ref{ebl_01505} that 
for the LIV scales we choose, $e^{-\tau}$ tends to one eventually, this is mainly because 
that the pair production process would be dramatically reduced or even forbidden due to the threshold anomalies, thus $e^{-\tau}$ tends to raise at higher energy beyond the threshold of the incident photon, and $e^{-\tau}$ then tends to approach one earlier for smaller LIV cases as the energy increases.

\begin{figure}[htbp]
    \centering
    \includegraphics[scale=0.53]{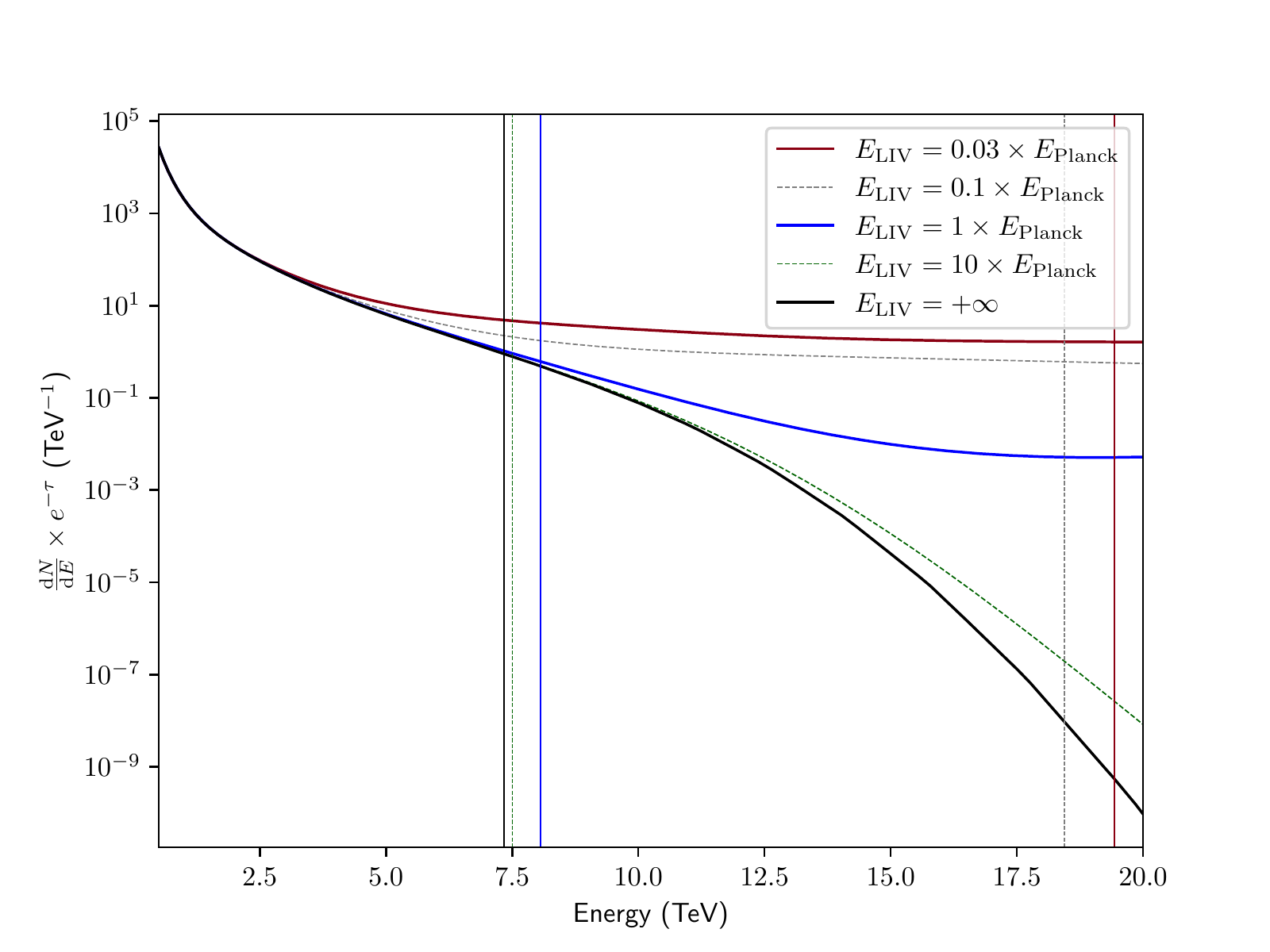}
    \caption{The hypothetical spectrum of GRB 221009A after EBL attenuation with different LIV scales.\label{spectrum_01505}}
\end{figure}

Next we combine the observation of GRB 221009A reported by LHAASO~\cite{LHAASO} and the EBL attenuation results in Fig.~\ref{ebl_01505} and calculate the potential effects of LIV numerically.
To do this, we first assume that the number density spectrum of GRB 221009A can be written as
\begin{equation}
    \frac{\textd N}{\textd E} = A_0\times E^{-\alpha},\label{spectrum}
\end{equation}
where the spectrum index $\alpha$ is chosen to be $2.5$ in this work~\footnote{\textadded{We choose \(\alpha=2.5\) in this work, while other reasonable values are possible.
Indeed LHAASO reported the time-dependent spectrum index for the energy band from \(\sim 200~\gev{}\) to \(\sim 7~\tev{}\) in Ref.~\cite{LHAASO:2023kyg}, according to which one can obtain the time-averaged index as \(2.41^{+0.14}_{-0.13}\), consistent with our choice of 2.5, especially when we take into account that if we extend the spectrum above 10~TeV, \(\alpha{}\) is expected to be larger than that of a lower energy band.
On the other hand, changing the value of \(\alpha{}\) would not affect the results of this work (at least qualitatively) as long as the value is proper~\cite{Li:2023rhj}, thus choosing 2.5 as a representative value of the index will not invalidate the analysis hereafter.}}, and $A_0$ is a constant to be normalized case by case.
The spectrum~\eqref{spectrum} we adopted is only assumed to be valid from 500~GeV to 20~TeV~\footnote{The upper limit of the energy range we choose is not 18~TeV since we take into consideration the uncertainty in the energy reconstruction 
of LHAASO.}, which is the energy range observed by LHAASO~\cite{LHAASO}.
Recalling that LHAASO observed more than 5000~photons\footnote{From now on we assume that there are 5000 photons observed in total.} from 500~GeV to 18(20)~TeV, we thus normalize the constant $A_0$ by the following strategy: multiplying $\textd N/\textd E$ by $e^{-\tau}$ first, where $\tau{}$ is either $\tau_{\gamma\gamma}$ or $\tau_\text{LIV}$ with the listed LIV scales, we then perform the integration below and equate the result to 5000,
\begin{equation}
    \int^{20~\tev}_{500~\gev}\textd E \frac{\textd N}{\textd E}(E)e^{-\tau(E,z=0.1505)}=5000,\label{normalization}
\end{equation}
from which we can determine $A_0$.
With $A_0$ determined for each case, we draw the spectra in Fig.~\ref{spectrum_01505}.
Just as we can see from Fig.~\ref{ebl_01505}, the larger the LIV scale is, the less the difference between the standard case and the LIV case.
To explain the observation of 18~TeV photons observed by LHAASO, we solve the following equation:
\begin{equation}
    \int^{20~\tev}_{E_\text{low}}\textd E\frac{\textd N}{\textd E}e^{-\tau(E,z)}=1~(\text{or }10^{-6})\label{number}
\end{equation}
for different LIV scales~\textadded{(including the standard case)}, where $E_\text{low}$ is the variable to be determined.
\textadded{For clarity some remarks are in order.
The left hand side of Eq.~\eqref{number} represents the {\em predicted\/} observed number of photons from GRB 221009A in the energy range from \(E_\text{low}\) to 20~TeV\@; the right hand side represents the photon number chosen to be the threshold signaling possible excesses. 
The choices of the threshold number are based on the following consideration: the choice of 1 is straightforward since for LHAASO to observe photons, at least one event is expected to reach the observatory, however the resulting \(E_\text{low}\) is merely an estimate because of the uncertainties from, for example, the distribution of EBL, and as an alternative we also consider a stricter threshold of \(10^{-6}\).
If we solve Eq.~\eqref{number} and obtain the value of \(E_\text{low}\), we then can compare it with 18~TeV.
Since the spectrum drops fast as the energy increases, the situation of \(E_\text{low}\ll 18~\tev{}\) clearly means that there is hardly any photon of around 18~TeV, leading to the conclusion that at lease we need to take into consideration the possibility of the existence of excesses.
The results of choosing 1 could serve as an alarm for the possible excess, while by choosing \(10^{-6}\) not only can we diminish the influence of uncertainties, we might obtain a stronger signal of excesses.
} 
\textadded{We also calculate the $E_{\text{low}}$ for observing 10 photons without LIV following the same logic, and the result reads \(\sim 4.67~\text{TeV}\). We think that this number could be taken as a rough estimate of the lower limit of the realistic energies of the top 10 energetic photons.}

The results for observing one photon are listed in Tab.~\ref{table1}.
From this table we can see that when $\eliv=\infty{}$, above 7.34~TeV we could observe one photon at least.
In Fig.~\ref{spectrum_01505} we show this results schematically.
In this figure, we also show the results in Tab.~\ref{table1} by 
vertical lines. 
As we can see, in the standard case, observing one photon around 18~TeV seems to be unlikely, while with a proper LIV scale this is permissible.
Indeed to observe more than $10^{-6}$ photon with $\eliv=\infty{}$, {\em i.e.}, within the framework of standard physics, we need $E_\text{low}\le 16.3~\tev{}$.
However this may contradict the report of observing around 18~TeV photons by LHAASO, and to observe at least one photon around 18~TeV, new physics, or a smaller index $\alpha<2.5$, or observed photon numbers more than 5000 is required.
In summary, the LHAASO result suggests 
the necessity 
to consider novel mechanisms to explain it, and LIV is 
a feasible candidate from our numerical results.

\begin{table}[htbp]
\caption{\label{table1}%
$E_\text{low}$ for observing one photon from GRB 221009A with different LIV scales.}
\begin{ruledtabular}
\begin{tabular}{cc}
\textrm{$E_\text{LIV}$~($\planck{}$)}&
\textrm{$E_\text{low}$ for observing one photon~(TeV)}\\
\colrule\
0.03 & 19.4 \\
0.1 & 18.4 \\
1 & 8.06 \\
10 & 7.51 \\
$\infty$ & 7.34\\
\end{tabular}
\end{ruledtabular}
\end{table}

\subsection{\label{generalgrb}General gamma-ray bursts}

The above result indicates the possibility of LIV with the observation of GRB 221009A, and in the following we discuss what we could learn from this for general GRBs, including short bursts and long bursts.
Some preparations are needed in order to ensure the comparability.
First, 
we choose the GRBs with the {\em best guess\/} values of the redshifts.
To be specific,  we assign $z=0.5$ 
for a typical short burst, while 
$z=2.15$ 
for a typical 
long burst~\cite{Huang2018}.
Second, we let the spectrum adopted hereafter be the same as the one in Eq.~\eqref{spectrum}, just like we put the same source to different locations in the universe.
However the energy range we aim to observe should be modified since we want to fix as many variables as possible.
Indeed, the energy range 500~GeV to 20~TeV corresponds to 575.25~GeV to 23.01~TeV at the source of GRB 221009A, as the observed energy $E_\text{obs}$ is related to the intrinsic one $E_\text{int}$ by
\begin{equation}
    \label{energyrelation}
    E_\text{int}=E_\text{obs}\times (1+z).
\end{equation}
We thus restrict ourselves to only analyze the corresponding energy ranges 383.5~GeV to 15.34~TeV and 182.6~GeV to 7.305~TeV for short bursts and long bursts respectively.
Third, 
let 
us assume that the GRB emits photons in an isotropic way, since the effective area of the observatory is fixed and the sources are identical, we have to consider the influence of distance.
As a consequence, photon numbers should be approximately multiplied by $(622.1~\text{Mpc}/1888.6~\text{Mpc}){}^2\approx 0.109$ and $(622.1~\text{Mpc}/5388.4~\text{Mpc}){}^2\approx 0.013$ respectively\footnote{The corresponding distances of $z=0.1505,0.5,2.15$ are calculated with the program at \url{https://www.astro.ucla.edu/~wright/CosmoCalc.html}, see also Ref.~\cite{Wright_2006}.}.

\begin{figure}
    \centering
    \includegraphics[scale=0.56]{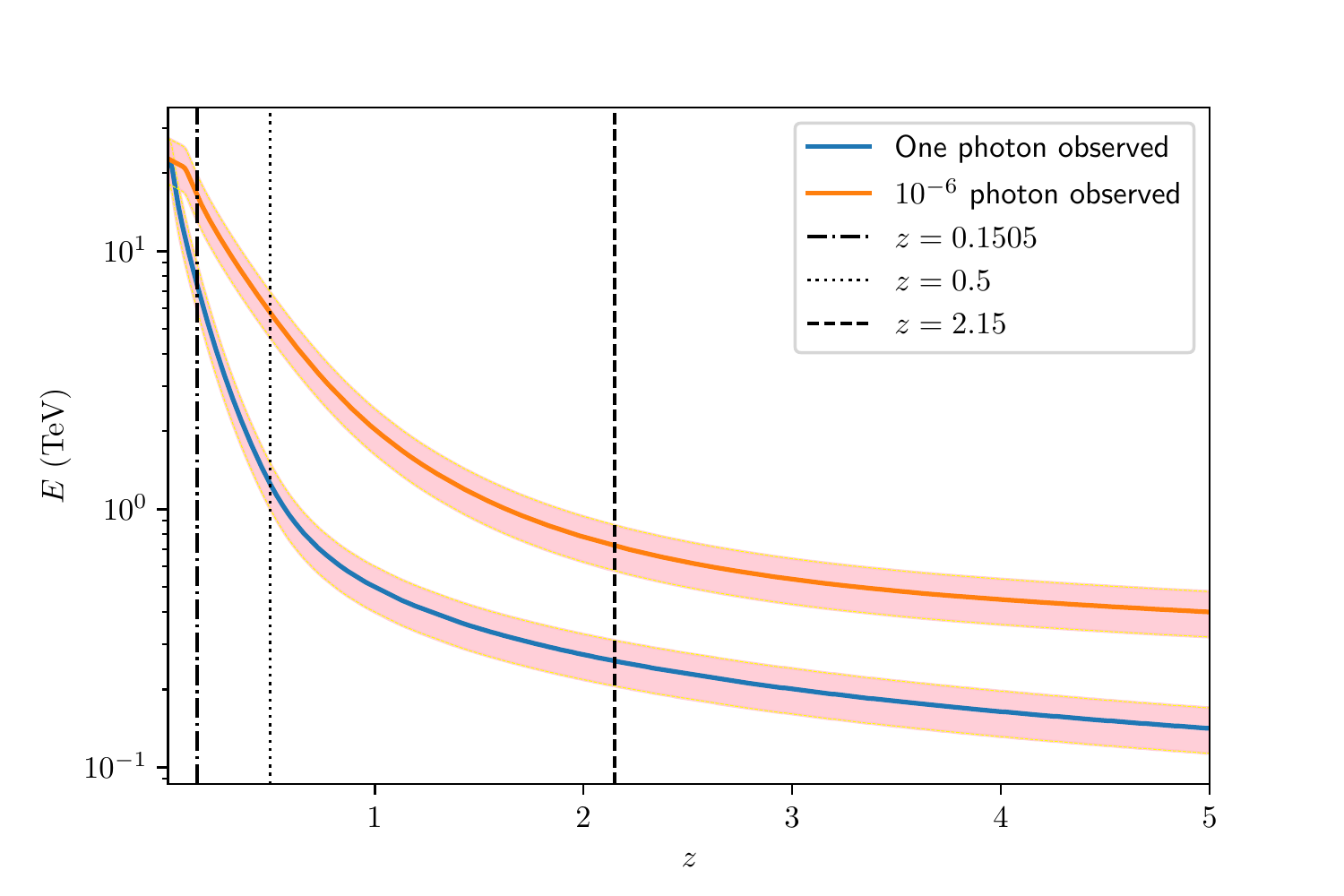}
    \caption{The lowest energies we can observe one photon~(and $10^{-6}$ photon) in the corresponding energy band from the same source as GRB 221009A at different redshifts.\label{energy_for_one}}
\end{figure}

Before analyzing GRBs with the {\em best guess\/} redshift values, we first perform a more general discussion:
we still fix the same conditions aforementioned, while now we let the redshift $z$ be a variable, and study above which energy we can observe one photon after EBL attenuation in the standard case by doing integration and solving the equation similar to Eq.~\ref{number} with the upper limit fixed but lower limit changing.
The result is shown in Fig.~\ref{energy_for_one} with the error estimated to be $20\%$.
From this figure, for example, we can see that if we observe a photon above 1~TeV from a GRB at $z=1$, then it could be considered as a starting point to seek for signal of LIV or other possibilities beyond the standard model.

In summary, we extend the energy band to the corresponding ones for short and long bursts, then do integration as in Eqs.~\eqref{normalization} and~\eqref{number} for different LIV scales.
Finally we multiply the results by 0.109 and 0.013 respectively.
As we will see, the result shows the uniqueness of GRB 221009A\@.

\subsubsection{General short bursts\label{shortbursts}}

\begin{figure}[htbp]
    \centering
    \includegraphics[scale=0.53]{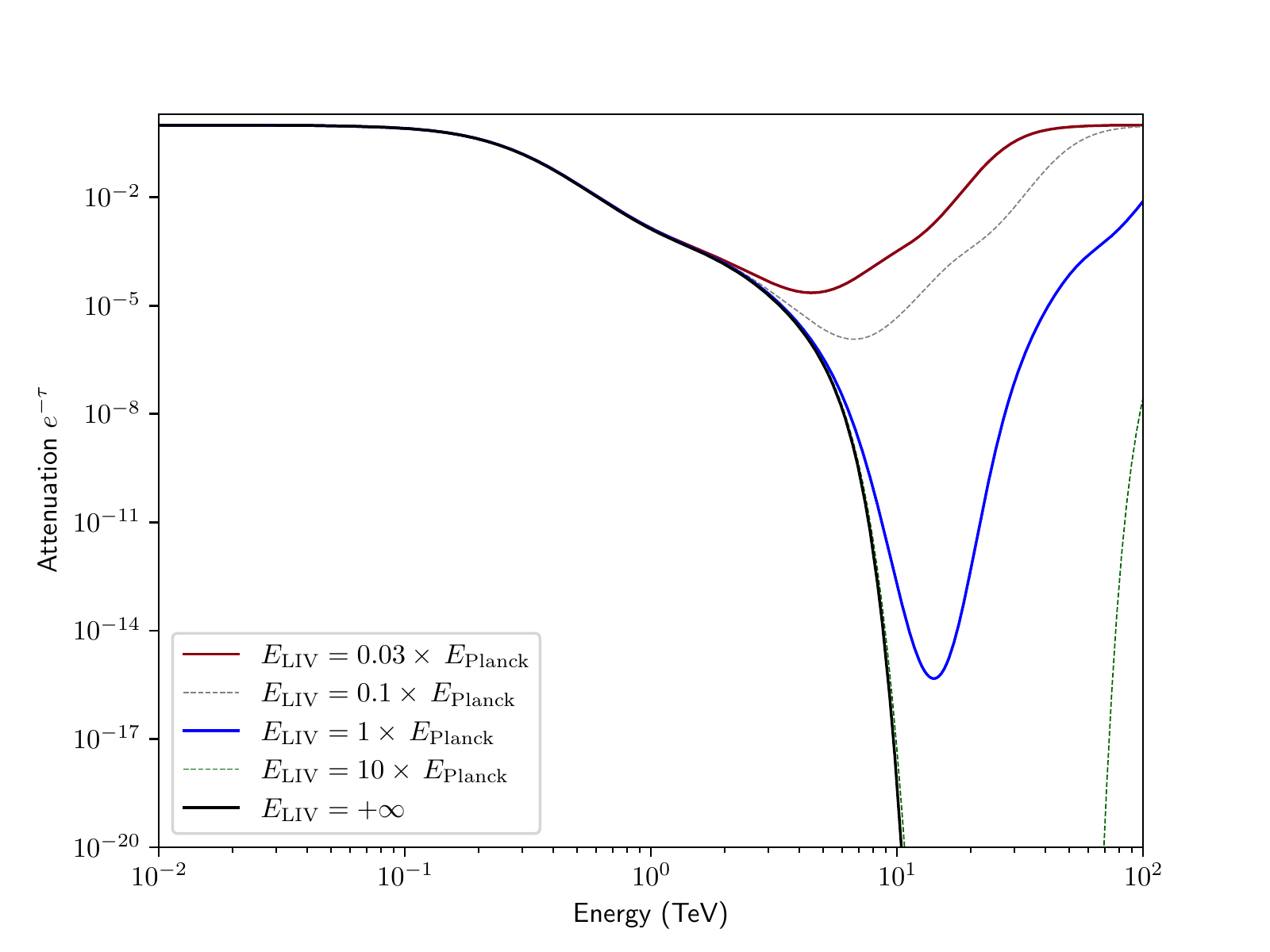}
    \caption{The EBL attenuation of $z=0.5$ with different LIV scales $\eliv{}$~\cite{Dominguez2011,manuel-meyer}\label{ebl_05}}
\end{figure}

\begin{figure}[htbp]
    \centering
    \includegraphics[scale=0.53]{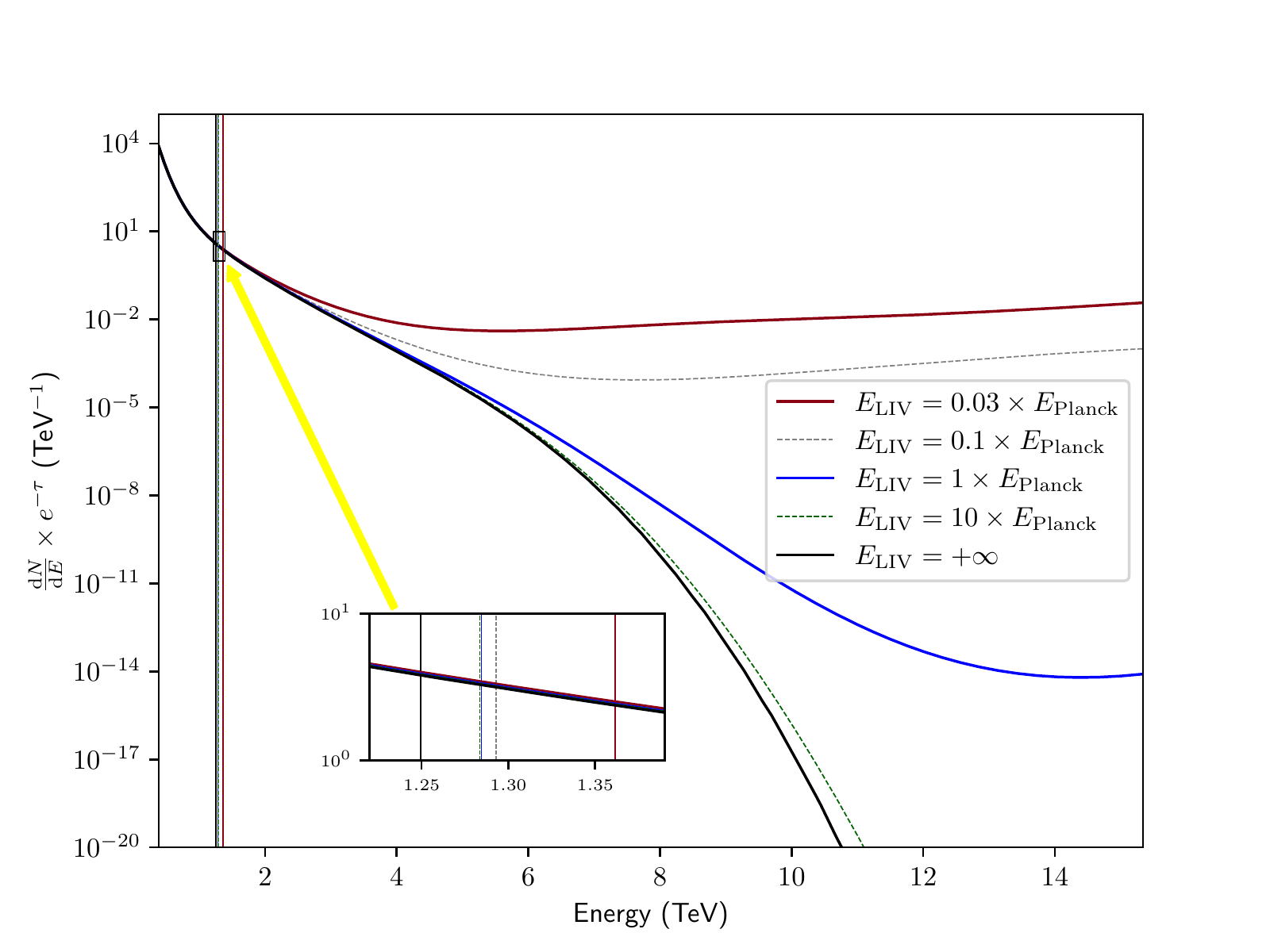}
    \caption{The hypothetical spectrum of a short GRB at $z=0.5$ after EBL attenuation with different LIV scales.\label{spectrum_05}}
\end{figure}

In Fig.~\ref{ebl_05}, we present the $E$-$e^{-\tau}$ curves~(produced by {\bfseries ebltable}~\cite{manuel-meyer}) for different LIV scales and the standard case 
as those
in Fig.~\ref{ebl_01505}, with the redshift set to be the {\em best guess\/} value $z=0.5$.
In this case we focus on the energy band $383.5~\gev{}$ to $15.34~\tev{}$, and it is shown in Fig.~\ref{ebl_05} that this band is affected by LIV as well, but the effects are relatively small compared to those in Fig.~\ref{ebl_01505}.
Fig.~\ref{spectrum_05} shows the spectra in this case.
In the standard case, to observe one photon we have $E_\text{low}\approx 1.25~\tev{}$~(the 
vertical line in Fig.~\ref{spectrum_05}) and to observe $10^{-6}$ photon we have $E_\text{low}\approx 5.76~\tev{}$.
However, the corresponding LIV results are close to the above results, as can be seen from the vertical lines in 
Fig.~\ref{spectrum_05}.  
This implies that it is hard to distinguish between different cases due to experimental uncertainties.
We will discuss this problem later. 

\subsubsection{General long bursts}

\begin{figure}[htbp]
    \centering
    \includegraphics[scale=0.53]{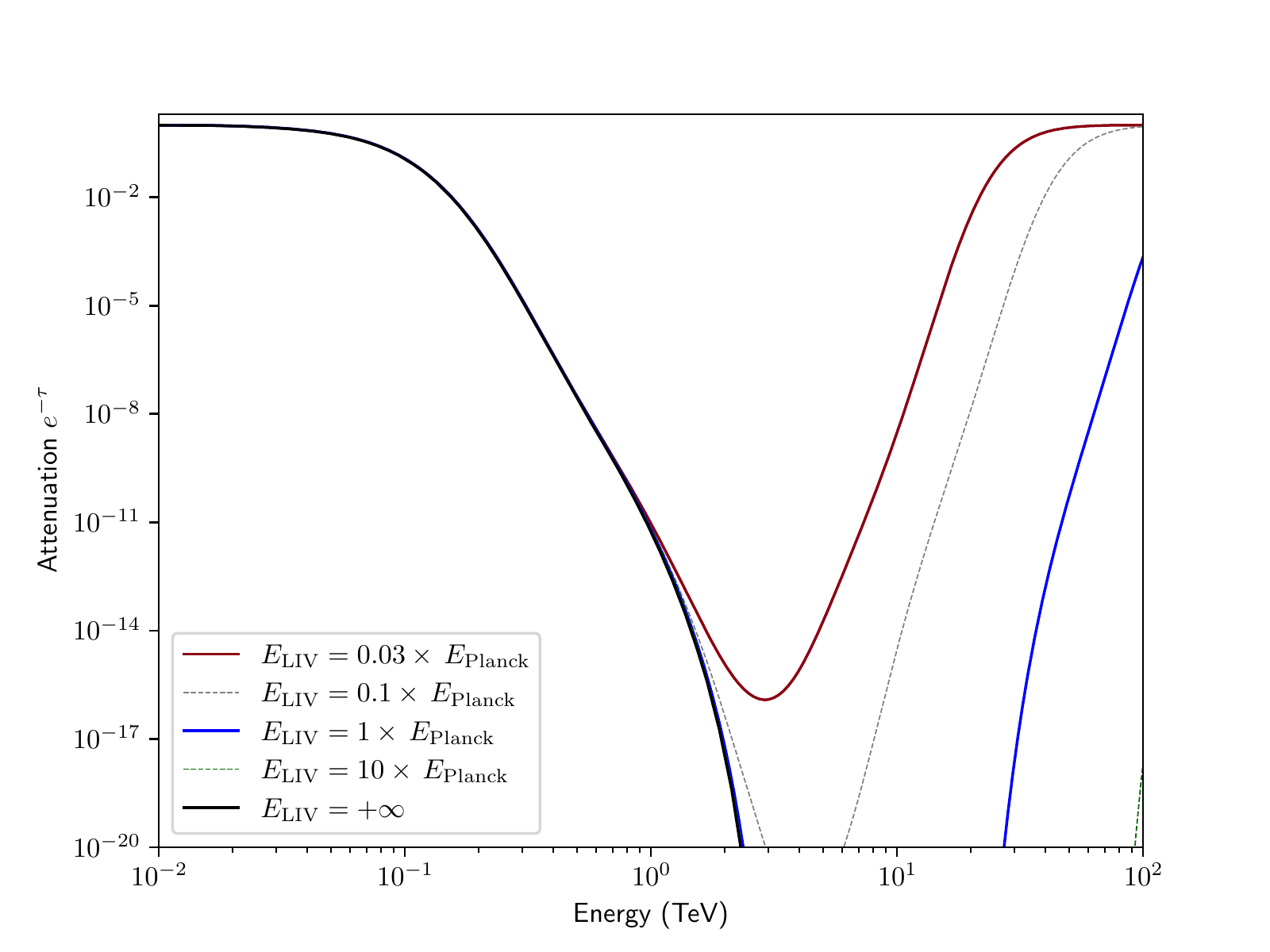}
    \caption{The EBL attenuation of $z=2.15$ with different LIV scales $\eliv{}$~\cite{Dominguez2011,manuel-meyer}\label{ebl_215}}
\end{figure}

\begin{figure}[htbp]
    \centering
    \includegraphics[scale=0.53]{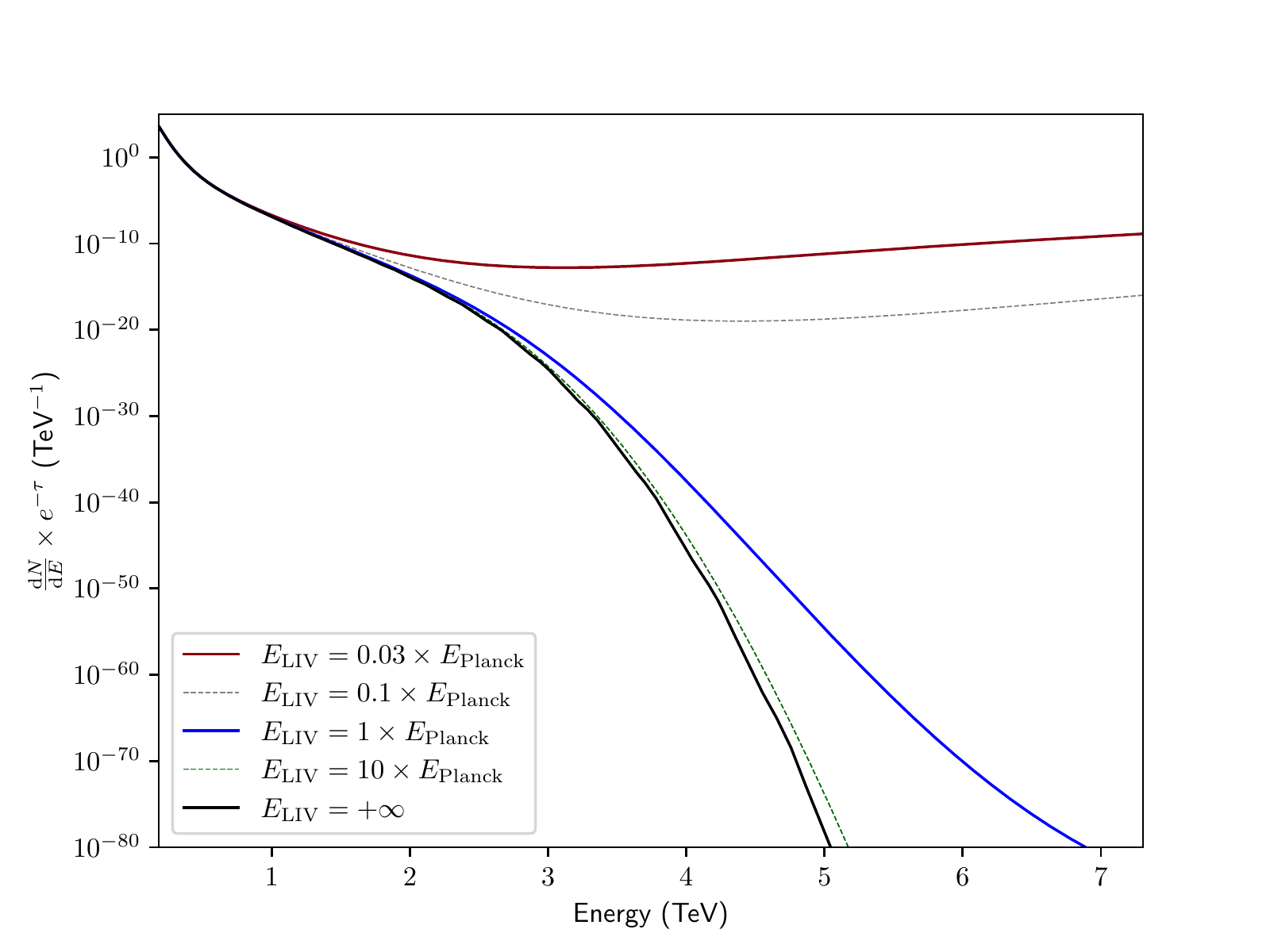}
    \caption{The hypothetical spectrum of a long GRB at $z=2.15$ after EBL attenuation with different LIV scales.\label{spectrum_215}}
\end{figure}

Similarly, using {\bfseries ebltable}~\cite{manuel-meyer} we draw the $E$-$e^{-\tau}$ diagram for a long GRB with $z=2.15$ in Fig.~\ref{ebl_215}, from which we can 
see
that even LIV is taken into consideration, the LIV effects are still too small to be distinguished between different cases, 
recalling that we focus on the energy band from 182.6~GeV to 7.305~GeV.
In fact, in Fig.~\ref{spectrum_215} we present the numerical results.
To observe one photon in the standard case, we have to require $E_\text{low}\ge 258~\gev{}$, while for $10^{-6}$ photon to be observed, $E_\text{low}\ge 720~\gev{}$.
Like the observation in the analysis for general short bursts, $E_\text{low}$ for observing one~(or $10^{-6}$) photon in each LIV case is not distinguishable enough from 
the standard case.
In other words, with the linear LIV term, a LIV scale $E_\text{LIV}$, which should have obvious indications in other approaches to LIV such as the study of time of light flight, is essential to make the standard case and LIV case distinguishable.
However this contradicts with the present phenomenological results of LIV\@.
Indeed from Fig.~\ref{spectrum_215} we may understand this problem easily.
As we can see, the spectra in Fig.~\ref{spectrum_215} all drop rapidly as the energy increase, so almost all photons fall into a narrow region from 182.6~GeV to about $300~\gev{}$.
Then a mild change in $E_\text{low}$ results in a dramatic change in the photon number, so the results for different cases are not distinguishable considering the experimental uncertainties of
observatories.
This fact is intriguing since this provides an approach to 
distinguish 
between LIV and other novel mechanisms such as axion-like particles.
This is because if there exists photon number excess and meanwhile $E_\text{low}$ for observation is obviously larger than that for the standard case, then both standard model and LIV are unlikely to be able to explain such observation.
So LIV is likely  
insensible to long GRBs and alternative interpretations should be 
sought.

Furthermore, the uniqueness of GRB 221009A for LIV and possibly other new physics proposals is clear from the analyses of general short and long bursts.
We find that for a reasonable LIV scale, there is 
less
chance that we can distinguish excess caused by LIV from the background or the standard results.
Nevertheless, this observation suggests that we can focus more on sources like GRB 221009A, with a relatively small redshift and very high energy messengers~(photons, neutrinos and electrons {\em etc.}), since this kind of objects is very unique for our studies of LIV and other new physics speculations.
We expect that with observatories such as LHAASO~\cite{Cao2010,DISCIASCIO2016166, Cao2019} and the Cherenkov Telescope Array~(CTA)~\cite{Actis2011,ACHARYA20133,Fairbairn2014,CTAConsortium:2017dvg}, more data will be available in the future.

\section{Discussions\label{discussions}}

In this work we re-analyzed the possibility of (linear) Lorentz invariance violation~(LIV) for photons as an explanation of the observation of very high energy photons from GRB 221009A reported by the Large High Altitude Air Shower Observatory~(LHAASO)~\cite{LHAASO}.
We also discussed if a gamma-ray burst~(GRB) the same as GRB 221009A is located at $z=0.5$ or $z=2.15$, the {\em best guess\/} value of the redshifts of short bursts or long bursts respectively~\cite{Huang2018}, to what extend our future observation can be influenced after fixing other conditions as possible.
We find that for GRB 221009A itself, a (linear) LIV scale $\eliv\lesssim\planck{}$ seems to be 
permissible 
with the observation by LHAASO\@.
While for a 
typical 
short or long burst, such a LIV scale is not enough 
and a mildly lower scale is 
sensible
for short bursts but we may not be able to observe very high energy photons from distant GRBs even LIV for photons does exist.
This shows the uniqueness of GRB 221009A, while we expect that observatories such as LHAASO and the Cherenkov Telescope Array~(CTA) will provide 
more data of sources similar to GRB 221009A, and then we can perform more detailed analyses of LIV\@.

Meanwhile, it is noteworthy that although LIV seems to be a 
candidate to render the observation of GRB 221009A reasonable, there are also other choices\textadded{, as we mentioned in Sec.~\ref{introduction}}.
\textadded{For these non-standard mechanisms, other phenomenological results are needed.
For example, if we want to confirm LIV, we could also utilize approaches such as time of light flight to study and search LIV~\cite{Amelino-Camelia1997a}; if we need to introduce new particles like heavy/sterile neutrinos~\cite{Cheung:2022luv,Smirnov:2022suv,Brdar:2022rhc,Huang:2022udc,Guo:2023bpo} or axion-like particles~(ALPs)~\cite{Galanti:2022pbg,Baktash:2022gnf,Troitsky:2022xso,Nakagawa:2022wwm,Zhang:2022zbm,Wang:2023okw}, we should be able to confirm the existence of these particles by conducting particle physics 
experiments at the same time.
As for the standard explanations, there are also accompanied signals.
It is clear that for the ultra-high-energy cosmic ray~(UHECR) mechanism to be established, one key characteristic is that the interaction between UHECRs and the background photons produces high energy neutrinos which could be recorded by observatories such as IceCube~\cite{AlvesBatista:2022kpg,Das:2022gon,Mirabal:2022ipw}.
Besides, the production mechanisms of UHECRs from gamma-ray bursts need to be established and work of comparing the predicted spectra with the observed ones is left to be finished.
In summary we can only conclude that all of these interpretations about the observations of GRB 221009A are promising equally, and at this stage we can hardly distinguish and clarify the origin of the observations of these above 10~TeV photons from GRB 221009A\@.
}
\textadded{We anticipate that by digging further into the present data of GRB 221009A from LHAASO and other observatories, future researches will shed light on the problems of whether there are non-standard mechanisms and if so, which mechanism is more favored.}

Before finishing this work, we should point out that we neglect the effects of the possible deviations in the cross-section~\eqref{livcrosssection}.
Indeed the effects of the extra terms may be important.
For example, if $\vartheta<0$ then the cross-section gets smaller and the linearity in $E$ makes it contribute more to the observation of high energy photons and as a result a larger $\eliv{}$ is permissible compared to the results in this work.
Otherwise a positive $\vartheta{}$ would imply a stronger LIV effect, thus require a smaller $\eliv{}$ 
to get the results for explaining the recent observation of GRB 221009A. This may break the available bound on $\eliv{}$ thus may render LIV less possible as the only
source for the recent observation of GRB 221009A and other interpretations are needed. 
But future work could still shed light on this question by further analyses in the future since the energy dependence of the change of the threshold and the extra terms in the cross-section 
are
quite different, so that we can constrain them separately with more data.

\begin{acknowledgments}
This work is supported by National Natural Science Foundation of China under Grants No.~12075003 and No.~12335006.
\end{acknowledgments}


\bibliography{ref}

\end{document}